\documentclass[traditabstract]{aa}
\usepackage{txfonts}
\usepackage{graphicx}
\usepackage{natbib}
\usepackage{supertabular}
\begin{document}

\title{Application of Bayesian model inadequacy criterion for multiple data sets to radial velocity models of exoplanet systems}
\author{Mikko Tuomi\inst{1,2}\thanks{The corresponding author, \email{m.tuomi@herts.ac.uk; mikko.tuomi@utu.fi}} \and David Pinfield\inst{1} \and Hugh R. A. Jones\inst{1}}
\institute{University of Hertfordshire, Centre for Astrophysics Research, Science and Technology Research Institute, College Lane, AL10 9AB, Hatfield, UK \and University of Turku, Tuorla Observatory, Department of Physics and Astronomy, V\"ais\"al\"antie 20, FI-21500, Piikki\"o, Finland}
\date{Received xx.xx.xxxx / Accepted xx.xx.xxxx}

\abstract
{We present a simple mathematical criterion for determining whether a given statistical model does not describe several independent sets of measurements, or data modes, adequately. We derive this criterion for two data sets and generalise it to several sets by using the Bayesian updating of the posterior probability density. To demonstrate the usage of the criterion, we apply it to observations of exoplanet host stars by re-analysing the radial velocities of HD 217107, Gliese 581, and $\upsilon$ Andromedae and show that the currently used models are not necessarily adequate in describing the properties of these measurements. We show that while the two data sets of Gliese 581 can be modelled reasonably well, the noise model of HD 217107 needs to be revised. We also reveal some biases in the radial velocities of $\upsilon$ Andromedae and report updated orbital parameters for the recently proposed 4-planet model. Because of the generality of our criterion, no assumptions are needed on the nature of the measurements, models, or model parameters. The method we propose can be applied to any astronomical problems, as well as outside the field of astronomy, because it is a simple consequence of the Bayes' rule of conditional probabilities.
}


\titlerunning{Bayesian model inadequacy criterion for multiple radial velocity data sets}

\maketitle


\section{Introduction}

Since the discovery of the first clear-cut example of an extrasolar planet orbiting a normal star \citep{mayor1995}, Doppler-spectroscopy, or radial velocity (RV), has been the most efficient method in detecting extrasolar planets orbiting nearby stars\footnote{See The Extrasolar Planets Encyclopaedia for an up-to-date list of known planetary candidates: \emph{http://exoplanet.eu/}.}.

Because the same nearby stars can be targets of several RV surveys, there is the possibility to combine the information of two or more RV data sets using the means of Bayesian inference \citep[e.g.][]{gregory2011,tuomi2011} and posterior updating. However, little is known about the possible biases individual data sets, or RV timeseries, may contain with respect to one another. Therefore, we use Bayesian tools in determining whether the common statistical models can be used to analyse RV timeseries without bias, and if not, how these models can be improved to receive trustworthy results. For these purposes we introduce a method for determining model inadequacy in describing multiple sets of measurements -- the Bayesian model inadequacy criterion.

The Bayes' rule leads naturally to the commonly used Bayesian model comparison methods \citep[e.g.][]{jeffreys1961}. These methods can be used efficiently to compare the relative performance of different statistical models of some \emph{a priori} selected model set. The Bayes' rule can be used to calculate the relative posterior probabilities of the models in the set given some measurements that describe some aspect of the modelled system. However, because only the relative performances of the models can be compared, it cannot be said whether the model with the greatest posterior probability is adequately accurate in describing the measured quantities.

The Bayes' factors \citep{kass1995,chib2001,ford2007} and other related measures of model goodness, such as the various information criteria \citep[e.g.][]{akaike1973,schwarz1978,spiegelhalter2002} derived using different approximations, can only be used to tell which one of the models in some model set describes the measurements the best -- i.e. the relative ``goodness`` of the models can be determined reliably. However, they cannot be used to assess whether this best model is as accurate description as possible given the information in the measurements. Our method of determining model inadequacy in this sense can be used to assess whether the model set can be estimated to contain a sufficiently accurate model that can be used to describe the measurements reliably.

Whether a given statistical model can be used to describe several sets of data in an adequate manner or not, has not been studied very extensively in the statistics literature. In \citet{kaasalainen2011}, the author presents a method for determining the optimal combination of two or more sources of data, or data modes. However, we are not aware of a single study discussing this problem in the Bayesian context, though \citet{spiegelhalter2002} appear to discuss the ''model adequacy'' in their article introducing the deviance information criterion, but they use the term interchangeably with the term ''fit''. Yet, determining whether a single model can describe two or more data sets without bias is of increasing importance in astronomy, particularly for indirect detections of the most interesting exoplanets whose signals lie close to the current limits of instrument sensitivity.

Since the RV variations of typical targets of Doppler-spectroscopy surveys are commonly modelled using a superposition of Keplerian signals, reference velocities, and possible linear trends, corrupted by some Gaussian noise, we use these models as a starting point of our analyses. However, we emphasise the fact that the RV variations caused by the stellar surface, usually referred to as the stellar jitter, are in general, despite some efforts in modelling their magnitude \citep{wright2005}, foreseen as arising from dark or bright spots primarily driven by stellar rotation \citep{barnes2011,boisse2011} and their effect on the RV's is not understood very well at the moment. Therefore, we model the excess noise in the RV's with care and show explicitly the statistical models we use in the analyses.

In section 2 we describe what we mean by the model inadequacy in describing two or more independent measurements or sets of measurements and provide a simple way of determining it in practice. We describe the details of our model inadequacy criterion in the Appendix. Finally, in section 3 we apply this criterion in practice by analysing astronomical RV exoplanet detections made using at least two different telescope-instrument combinations.

\section{Bayesian analyses and model inadequacy}

The Bayesian methods do not differentiate between determining the most probable parameter values or most probable models containing these parameters. They can all be arranged into a linear order, which yields information on the observed system if only the selected models describe the observed system realistically enough. It is possible to calculate the relative posterior probabilities of any number of models and determine their relative magnitudes in a similar way as it is possible to determine the posterior odds of having the measurements drawn from a probability density characterised by a certain parameter value of any one of the models. We do not describe the process of determining the posterior probability densities of the model parameters here, because several well-known posterior sampling methods exist and they have been well covered by the existing literature \citep[e.g.][]{metropolis1953,hastings1970,geman1984,haario2001}. The performance of these methods has also been demonstrated by several re-analyses of existing RV data, revealing the existence of planets \citep[e.g.][]{gregory2005,gregory2007a,gregory2007b,tuomi2009} or disputing it \citep[e.g.][]{tuomi2011}. In these works, the model probabilities have played an important role in assessing the number of planetary companions orbiting nearby stars.

Commonly, the Bayesian tools are used to assess the probabilities of different statistical models given the measurements $m$ that are being analysed using the models. These tools provide the relative probabilities of the selected models $\mathcal{M}_{i}, i=1, ..., k$ in the \emph{a priori} determined model set as
\begin{equation}\label{model_probability}
  P(\mathcal{M}_{i} | m) = \frac{P(m | \mathcal{M}_{i})P(\mathcal{M}_{i})}{\sum_{j=1}^{k}P(m | \mathcal{M}_{j})P(\mathcal{M}_{j})} ,
\end{equation}
where probabilities $P(\mathcal{M}_{i}), i=1, ..., k$ are the prior probabilities of the different models and the marginal likelihoods $P(m | \mathcal{M}_{i})$ are defined as
\begin{equation}\label{marginal_likelihood}
  P(m | \mathcal{M}_{i}) = \int_{\theta_{i} \in \Theta_{i}} l(m | \theta_{i}) \pi(\theta_{i}) d \theta_{i} ,
\end{equation}
where $\pi(\theta_{i})$ is the prior probability density of the parameter or parameter vector $\theta_{i}$ of the model $\mathcal{M}_{i}$ and $l(m | \theta_{i})$ represents the likelihood function corresponding to the model.

The interpretation of the posterior probabilities in Eq. (\ref{model_probability}) is a rather subjective matter because they are relative and it is only possible to assess how much confidence one has in one of the models compared to the rest of them. According to the views of \citet{jeffreys1961,kass1995}, a model would have to be at least 150 times more probable than the next best model to have strong evidence in favour of it. We adopt the same threshold because claiming that there are $k+1$ planets orbiting a star instead of $k$ needs to be on a solid ground with respect to the model probabilities. Especially, if the model with $k+1$ planets was e.g. 50 times more probable than that with $k$ planets, there would still be a roughly 2\% possibility that the $k$ planet model explains the data. Therefore, we choose a rather high threshold when interpreting the posterior probabilities of models with different numbers of Keplerian signals.

With the marginal likelihoods available according to the Eq. \ref{marginal_likelihood}, we define the model $\mathcal{M}$ to be an inadequate description of independent measurements $m_{i}, i=1, ..., N$, if it holds that for some small positive number $r$
\begin{equation}\label{MIC}
  B(m_{1}, ..., m_{N} | \mathcal{M}) : = \frac{P(m_{1}, ..., m_{N} | \mathcal{M})}{\prod_{i=1}^{N} P(m_{i} | \mathcal{M})} < r .
\end{equation}
This definition is based on the independence of the measurements and that they are being modelled with a single statistical model. It is a simple result of a relation of the marginal likelihoods of each of the measurement and the joint marginal likelihood of all of them shown in Eq. (\ref{multiple_probabilities}). We derive this criterion using the Bayes' rule of conditional probabilities and the concept of independence, and also interpret the results in terms of information theory in the Appendix.

The number $r$ has an interpretation as a threshold value. For instance, the model being inadequate with probabilities 90\%, 95\%, and 99\% corresponds to threshold values of 0.111, 0.053, and 0.010, respectively (see Appendix). Therefore, if the best model according to Eq. (\ref{model_probability}) satisfies Eq. (\ref{MIC}) for some reasonably small $r$, it can be concluded that the model does not describe the measurements without bias and the corresponding analysis results may be biased as well. In such a case, the model set has to be re-considered and expanded by adding better descriptions of the data to it. In practice, we use the 95\% threshold value, but choosing its value is a subjective issue and only represents how confidently one wants to determine the model inadequacy.

We note that the model inadequacy can also be interpreted in terms of the measurements being inconsistent with one another with respect to the model used. This interpretation arises from the fact that the model may not take into account some features in one or more data sets that result from biases in the process of making the measurements or from some other unmodelled features in the data. We use the inadequacy of the model given the data sets and the inconsistency of the data sets with respect to this model interchangeably throughout this article.

We describe the parameter probability densities using three numbers. These numbers are the maximum \emph{a posteriori} (MAP) estimate of the posterior density and the limits of the 99\% Bayesian credibility set $\mathcal{D}_{0.99}$ as defined in e.g. \citet{tuomi2009}. We calculate these estimates from the posterior densities of the model parameters received using the adaptive Metropolis posterior sampling algorithm \citep{haario2001}, which is a modification of the famous Metropolis-Hastings (M-H) algorithm \citep{metropolis1953,hastings1970} that adapts the proposal density to the shape of the posterior density of the model parameters. Because of this property, it is not very sensitive to the choise of initial parameter vector nor proposal density -- desired features that make the method significantly more robust than the common M-H algorithm by enabling a more rapid convergence to the posterior.

While the adaptive Metropolis algorithm assumes a Gaussian proposal density, it adapts to the posterior reasonably rapidly and a samples of roughly $10^{6}$ are sufficient for the chain burn-in period in all the analyses, i.e. until the chain converges to the posterior. Because of the Gaussian posterior, the acceptance rate of the chain can sometimes decrease to as low values as 1\% when the posterior density is highly nonlinear, as is comonly the case with RV data. However, in such cases, we simply increased the chain length by a factor of 10-20 and saved computer memory by only saving every 10th or 20th member of the chain to the output file. We verified that the chain had indeed converged by running up to five samplings with different initial values and required that they all produced marginal integrals that were equal up to the second digit. With a converged chain, we then calculated the marginal likelihoods using the method of \citet{chib2001}. 

For the sake of trustworthiness, throughout this article we also take into account the uncertainties in the stellar masses when calculating the semi-major axes and RV masses of the planets orbiting them. These uncertainties are taken into account by using a direct Monte Carlo simulation -- i.e. by drawing random values from both the density of the model parameters and the estimated density of the stellar mass when calculating the densities of the semi-major axes and planetary RV masses. We assume that the estimated distribution of the stellar mass, usually reported using mean and stardard error, is independent of the densities of the orbital parameters from the posterior samplings.

\section{Model inadequacy criterion and exoplanet detections}

In principle, analysing RV data is reasonably simple because the planet induced stellar wobble can be modelled using the well-known Newtonian laws of motion -- especially if the gravitational planet-planet interactions are not significant in the timescale of the observations and post-Newtonian effects are negligible. In practice, though, there are several aspects of the RV measurements that are not understood well enough to be able to consider that the models describe the measurements in an adequate manner. These aspects include e.g. disturbances caused by undetected planets or planets whose orbital periods cannot be constrained \citep[e.g.][]{ford2007}; noise caused by the inhomogeneities in the stellar surface, usually referred to as the stellar ''jitter'' \citep[e.g.][]{wright2005}; and excess noise and possible biases that are particular to the various instruments and telescopes used to make the observations. All these aspects make the analyses of RV's challenging and if not accounted for properly by the statistical models used, can lead to biased results and misleading interpretations.

In this section we re-analyse three RV data sets made using at least two telescope-instrument combinations. Assuming these sets are independent -- which is a common assumption, though not explicitly stated most of the time when analysing several sets of measurements -- we apply the model inadequacy criterion to find out if the common models should be modified and if the corresponding results are different from the ones found in the literature.

\subsection{HD 217107}

The RV's of HD 217107 are known to contain the signatures of two extrasolar planets \citep{fischer1999,fischer2001,vogt2000,naef2001,vogt2005,wittenmyer2007,wright2009}. The system consists of a massive short-period planet with an orbital period of roughly 7 days, and an outer long-period planet with an orbital period of 11 years. The RV's of this target have been observed using 4 instruments mounted on 5 telescopes, namely, Euler \citep{naef2001}, Harlam J. Smith (HJS) \citep{wittenmyer2007}, Keck I \citep{wright2009}, and Shane and Coude Auxiliary Telescope (CAT) at the Lick Observatory \citep{wright2009}. Together, there are 293 RV measurements of this system.

The most up-to-date solution is that of \citet{wright2009}, where the combined Keck and Lick data with 207 measurements was analysed. However, the authors do not discuss the exact statistical model used in their analyses and therefore we feel that this combined data set should be re-analysed to see how the four data sets should be modelled to receive the most trustworthy results.

Following the common Bayesian approach \citep[e.g.][]{gregory2005,gregory2007a,gregory2007b,tuomi2009,tuomi2011}, we choose our model set to consist of four models, namely, models $\mathcal{M}_{k}, k = 0, ..., 3$, where $k$ denotes the number of planetary signals in the data. Therefore, there are $5k + 5$ parameters in our models corresponding to 5 parameters for each planet -- RV amplitude $K$, orbital eccentricity $e$, orbital period $P$, longitude of pericentre $\omega$, and  mean anomaly $M_{0}$, i.e. the date of periastron passage as expressed in radians between 0 and 2$\pi$ -- four parameters decribing the reference velocities of each data set $\gamma_{l}, l = 1, ..., 4$, and the parameter describing the magnitude of stellar jitter $\sigma_{J}$. Our set of statistical models describing the measurement $m_{i,l}$ made at time $t_{i}$ is
\begin{equation}\label{RV_model}
  r_{k}(t_{i}) + \gamma_{l} + \epsilon_{i} + \epsilon_{J} , k=0, ..., 3 ,
\end{equation}
where $r_{k}$ represents the $k$ Keplerian signals and $\epsilon_{i}$ and $\epsilon_{J}$ are Gaussian random variables with zero mean and known variance $\sigma_{i}^{2}$ and an unknown variance $\sigma_{J}^{2}$, respectively. The variance $\sigma_{i}^{2}$ corresponds to the instrument uncertainty of each individual measurement, which is usually assumed known and is reported together with the data.

We analyse the combined data set using the models $\mathcal{M}_{k}, k = 0, ..., 3$ and receive the model probabilities in Table \ref{HD217107_probabilities}. These probabilities imply that there are two companions orbiting the star with high confidence. However, the Bayes factor determining the inadequacy of the best model in the model set has to be calculated to assess the reliability of this model. Denoting the four data sets as $m_{l}, l = 1, ..., 4$, we receive $B(m_{1}, ..., m_{4}) = 0.05$, which means that the model is an inadequate description of the data with a probability of 0.95. This implies, that the model set does not contain a sufficiently good model, i.e. the data sets are not consistent with one another given this model, and needs to be expanded.

\begin{table}
\center
\caption{\label{HD217107_probabilities}The relative model probabilities of $k$ planet models for the combined data set of HD 217107.}
\begin{tabular}{lc}
\hline \hline
  $k$ & $P(\mathcal{M}_{k})$ \\
\hline
  0 & $< 10^{-272}$ \\
  1 & $< 10^{-85}$ \\
  2 & 1.00 \\
  3 & $< 10^{-2}$ \\
\hline \hline
\end{tabular}
\end{table}

Because of the inadequacy of model $\mathcal{M}_{2}$, we no longer assume that the instrument noise is known according to the variances $\sigma_{i}^{2}$ but suspect that there could be unknown random variations or biases that differ between the data sets. Therefore, we expand our model set by models
\begin{equation}\label{RV_model2}
  r_{k}(t_{i}) + \gamma_{l} + \epsilon_{i} + \epsilon_{I,l} , k=1, .., 3,
\end{equation}
where the Gaussian random variable $\epsilon_{I,l}$ is different for every data set and is assumed to consist of additional random variation caused by the instrument noise and stellar jitter. Therefore, in this model, the resulting values $\sigma_{I,l}$ can only be interpreted as giving the upper limit for the stellar jitter. We denote these models as $\mathcal{M}_{I,k}$.

Using the expanded model set, we receive the model probabilities in Table \ref{HD217107_probabilities2}. These probabilities imply that there are indeed differences in the noise levels of the different data sets and that these differences have to be taken into account when assessing the orbital parameters of the planets. We calculate the model inadequacy Bayes factor $B(m_{1}, ..., m_{4})$ for the best model $\mathcal{M}_{I,2}$. This time $B(m_{1}, ..., m_{4}) = 3.3 \times 10^{12}$, which corresponds to an inadequacy probability of $3.0 \times 10^{-13}$, a value that clearly states the best model cannot be considered inadequate.

\begin{table}
\center
\caption{The relative model probabilities of $k$ planet models $\mathcal{M}_{k}$ and $\mathcal{M}_{I,k}$ for the combined data set of HD 217107 (all 8 probabilities are on the same scale).\label{HD217107_probabilities2}}
\begin{tabular}{lcc}
\hline \hline
  $k$ & $P(\mathcal{M}_{k})$ & $P(\mathcal{M}_{I,k})$ \\
\hline
  0 & $< 10^{-286}$ & $< 10^{-289}$ \\
  1 & $< 10^{-99}$ & $< 10^{-96}$ \\
  2 & $< 10^{-14}$ & 1.00 \\
  3 & $< 10^{-17}$ & $< 10^{-2}$ \\
\hline \hline
\end{tabular}
\end{table}

\begin{figure}
\center
\includegraphics[angle=270, width=0.45\textwidth]{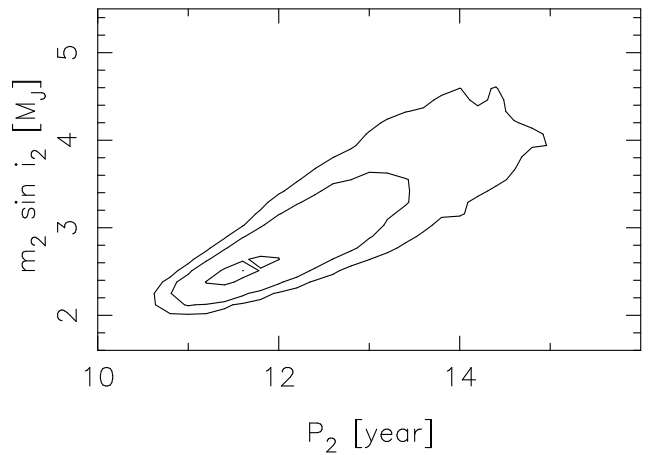}
\caption{The equiprobability contours of the RV mass and orbital period of HD 217107 c containing 50\%, 95\%, and 99\% of the probability density.}\label{contour_HD217107}
\end{figure}

We have listed the solution of the model with the greatest posterior probability, $\mathcal{M}_{I,2}$, in Table \ref{HD217107_parameters}. While consistent with the results of \citet{wright2009}, our solution with the best model $\mathcal{M}_{I,2}$ has much more uncertain parameter values, especially for the period, RV mass, and RV amplitude of the outer companion, which is also found heavily correlated with the reference velocity parameters. We show the 99\%, 95\%, and 50\% equiprobability contours of RV mass and period of the outer companion in Fig. \ref{contour_HD217107} (the gap in the 50\% contours arises from the numerical inaccuracy of the plot). This Fig. is similar to the Fig. 8 in \citet{wright2009}, but they used the $\chi^{2}$ density for the plot instead of posterior density. Also, we note that the jitter of HD 217107 has a level of at most 6.0 ms$^{-1}$ based on the noise in the Euler data, which turned out to contain the least noise out of the four data sets. It is also interesting to see that the Lick data had therefore at least 5 ms$^{-1}$, but possibly even more than 10.0 ms$^{-1}$, additional uncertainty that can only be caused by the telescopes and the instrument. Therefore, it cannot be said that the Lick instrument uncertainty is known according to the standard uncertainties of the data reduction pipeline, as reported when publishing Lick RV's. This could in fact be one of the reasons the parameter values in our solution (Table \ref{HD217107_parameters}) appear to be more uncertain than those reported by \citet{wright2009}, though they do not indicate the confidence-level of the reported uncertainties.

\begin{table*}
\center
\caption{The two-planet solution of HD 217107 combined data set. The MAP estimates of the parameters and in brackets the limits of their $\mathcal{D}_{0.99}$ sets. The solution of \citet{wright2009} is shown for comparison for the corresponding parameters as reported by them.\label{HD217107_parameters}}
\begin{tabular}{lcccc}
\hline \hline
Parameter & $\mathcal{M}_{I,2}$ & $\mathcal{M}_{I,2}$ & \multicolumn{2}{c}{\citet{wright2009}} \\
& Planet b & Planet c & Planet b & Planet c \\
\hline
$P$ [days] & 7.12664 [7.12674, 7.12692] & 4300 [3800, 6000] & 7.126816(39) & 4270(220) \\
$e$ & 0.123 [0.111, 0.139] & 0.49 [0.39, 0.58] & 0.1267(52) & 0.517(33) \\
$K$ [ms$^{-1}$] & 138.3 [136.0, 140.1] & 31.5 [25.0, 60.4] & 139.20(92) & 35.7(1.3) \\
$\omega$ [rad] & 0.39 [0.29, 0.52] & 3.38 [3.12, 3.82] & & \\
$M_{0}$ [rad]& 4.97 [4.85, 5.08] & 1.44 [0.63, 1.80] & & \\
$m_{p} \sin i$ [M$_{\oplus}$] & 1.35 [1.22, 1.47] & 2.6 [1.8, 5.4] & 1.39(11) & 2.60(15) \\
$a$ [AU] & 0.0742 [0.0701, 0.0771] & 5.3 [4.7, 6.6] & 0.0748(43) & 5.32(38) \\
\hline
$\gamma_{1}$ [ms$^{-1}$] (Euler) & 6.6 [-12.7, 14.2] \\
$\gamma_{2}$ [ms$^{-1}$] (HJS) & 11.0 [-6.9, 19.0] \\
$\gamma_{3}$ [ms$^{-1}$] (Keck) & -0.8 [-19.9, 4.9] \\
$\gamma_{4}$ [ms$^{-1}$] (Lick) & -1.2 [-19.9, 5.0] \\
$\sigma_{I,1}$ [ms$^{-1}$] (Euler) & 2.7 [0.0, 6.0] \\
$\sigma_{I,2}$ [ms$^{-1}$] (HJS) & 4.8 [1.1, 8.4] \\
$\sigma_{I,3}$ [ms$^{-1}$] (Keck) & 5.4 [4.4, 6.4] \\
$\sigma_{I,4}$ [ms$^{-1}$] (Lick) & 12.9 [10.9, 15.4] \\
\hline \hline
\end{tabular}
\end{table*}

\subsection{Gliese 581}

The Gliese 581 planetary system has been claimed to be a host to as many as six relatively low-mass planets \citep{bonfils2005,udry2007,mayor2009,vogt2010}. Though the most likely number of planetary companions in the system is four \citep{tuomi2011} or five \citep{gregory2011}, the RV's of Gliese 581 provide a challenging analysis problem because the signals are only barely distinguishable from the relatively noisy measurements.

We start by analysing the combined data set of HARPS and HIRES RV measurements \citep[see e.g.][]{vogt2010,gregory2011,tuomi2011} using the models $\mathcal{M}_{k}$ and $\mathcal{M}_{I,k}$ with $k = 0, ..., 5$. We choose this model set because we already suspect, based on the analysis of the RV's of HD 217107, that this combined data set may have different noise levels corresponding to the different telescope-instrument combinations.

The posterior probabilities of the models in our model set are shown in Table \ref{GJ581_probabilities}. These probabilities, while having the greatest value for model $\mathcal{M}_{I,5}$, do not support the conclusion that there are five Keplerian signals in the data strongly enough because the probability of model $\mathcal{M}_{I,4}$ is highly significant. Therefore, we check the inadequacy of the latter model to see if our statistical model is good enough.

\begin{table}
\center
\caption{The relative model probabilities of $k$ planet models $\mathcal{M}_{k}$ and $\mathcal{M}_{I,k}$ for the combined data set of Gliese 581.\label{GJ581_probabilities}}
\begin{tabular}{lcc}
\hline \hline
  $k$ & $P(\mathcal{M}_{k})$ & $P(\mathcal{M}_{I,k})$ \\
\hline
  0 & $< 10^{-128}$ & $< 10^{-129}$ \\
  1 & $< 10^{-33}$ & $< 10^{-34}$ \\
  2 & $<10^{-13}$ & $< 10^{-14}$ \\
  3 & $<10^{-10}$ & $< 10^{-10}$ \\
  4 & $<10^{-2}$ & 0.16 \\
  5 & 0.11 & 0.72 \\
  6 & $<10^{-2}$ & $<10^{-2}$ \\
\hline \hline
\end{tabular}
\end{table}

The Bayes factor in Eq. (\ref{MIC}) has a value of $2.0 \times 10^{10}$ for the four-planet model $\mathcal{M}_{I,4}$, which means that the probability of the HIRES and HARPS data sets being inadequately described by the model is $5.0 \times 10^{-11}$, a value low enough to conclude that there is no need to revise the model. We note that this model, an order of magnitude more probable than the previously used model $\mathcal{M}_{4}$ \citep{tuomi2011}, does not result in a revision of the orbital parameters (Table \ref{Gliese_parameters}). However, the noise parameters of the two data sets do differ from one another slightly. Denoting the HIRES data set with $l=1$ and the HARPS data set with $l=2$, the parameters $\sigma_{I,l}, l=1,2$, have MAP estimates of 2.39 and 1.50 ms$^{-1}$, respectively. The corresponding 99\% credibility sets are [1.77, 3.09] and [1.00, 2.01] ms$^{-1}$, respectively. Therefore, the noise in the HARPS measurements gives an upper limit for the jitter of Gliese 581 of 2.01 ms$^{-1}$, whereas there is likely a small amount of additional instrument noise in the HIRES data.

\begin{table*}
\center
\caption{The four-planet solution of GJ 581 combined HARPS and HIRES data. The MAP estimates of the parameters and in brackets the limits of their $\mathcal{D}_{0.99}$ sets.\label{Gliese_parameters}}
\begin{tabular}{lcccc}
\hline \hline
Parameter & Planet e & Planet b & Planet c & Planet d \\
\hline
$P$ [days] & 3.1487 [3.1479, 3.1507] & 5.36845 [5.36810, 5.36890] & 12.917 [12.908, 12.926] & 66.88 [66.12, 67.32] \\
$e$ & 0.05 [0, 0.38] & 0.005 [0, 0.048] & 0.04 [0, 0.24] & 0.36 [0, 0.65] \\
$K$ [ms$^{-1}$] & 1.76 [1.08, 2.37] & 12.45 [11.90, 13.07] & 3.26 [2.67, 3.92] & 1.83 [1.15, 2.52] \\
$\omega$ [rad] & 2.4 [0, 2$\pi$] & 3.9 [0, 2$\pi$] & 2.6 [0, 2$\pi$] & 5.6 [0, 2$\pi$] \\
$M_{0}$ [rad] & 2.6 [0, 2$\pi$] & 2.6 [0, 2$\pi$] & 3.5 [0, 2$\pi$] & 4.7 [0, 2$\pi$] \\
$m_{p} \sin i$ [M$_{\oplus}$] & 1.86 [1.14, 2.51] & 15.73 [14.38, 16.95] & 5.51 [4.45, 6.56] & 5.19 [3.36, 7.21] \\
$a$ [AU] & 0.0284 [0.0275, 0.0294] & 0.0406 [0.0393, 0.0420] & 0.0728 [0.0706, 0.0751] & 0.218 [0.211, 0.226] \\
\hline
$\gamma_{1}$ [ms$^{-1}$] (HARPS) & -0.36 [-0.88, 0.12] \\
$\gamma_{2}$ [ms$^{-1}$] (HIRES) & 0.38 [-0.41, 1.17] \\
$\sigma_{I,1}$ [ms$^{-1}$] (HARPS) & 1.50 [1.00, 2.01] \\
$\sigma_{I,2}$ [ms$^{-1}$] (HIRES) & 2.39 [1.77, 3.09] \\
\hline \hline
\end{tabular}
\end{table*}

\subsection{$\upsilon$ Andromedae}

The RV's of $\upsilon$ And have shown three strong Keplerian signals resulting from three massive planets orbiting the star \citep{butler1997,butler1999,fischer2003,naef2004,wittenmyer2007,wright2009}. The star has been a target of five RV surveys for several years, namely, Lick \citep{butler1999,fischer2003,wright2009}, the Advanced Fiber-Optic Echelle spectrometer (AFOE) at the Whipple Observatory \citep{butler1999}, HJS \citep{wittenmyer2007}, ELODIE at the Haute-Provence Observatory \citep{naef2004}, and the Hobby-Eberly Telescope (HET) \citep{mcarthur2010}. Recently, the combined data of Lick \citep{fischer2003,wright2009} and ELODIE \citep{naef2004} has been reported to contain a fourth planetary signal \citep{curiel2011}.

We re-analyse the combined RV data of $\upsilon$ And by using the model inadequacy criterion. However, before we start, we check the consistency of the 248 Lick RV's published in \citet{fischer2003} and the 284 Lick RV's published in \citet{wright2009} (we denote these data sets as Lick1 and Lick2, respectively), because \citet{curiel2011} used Lick2 data and the additional 30 RV points from Lick1 that were not included in Lick2. The fact that these 30 measurements were not included in Lick2 likely because of suspected biases or calibration errors suggests that there could be some biases within the combined Lick data analysed in \citet{curiel2011} as well.

\begin{figure}
\center
\includegraphics[angle=270, width=0.45\textwidth]{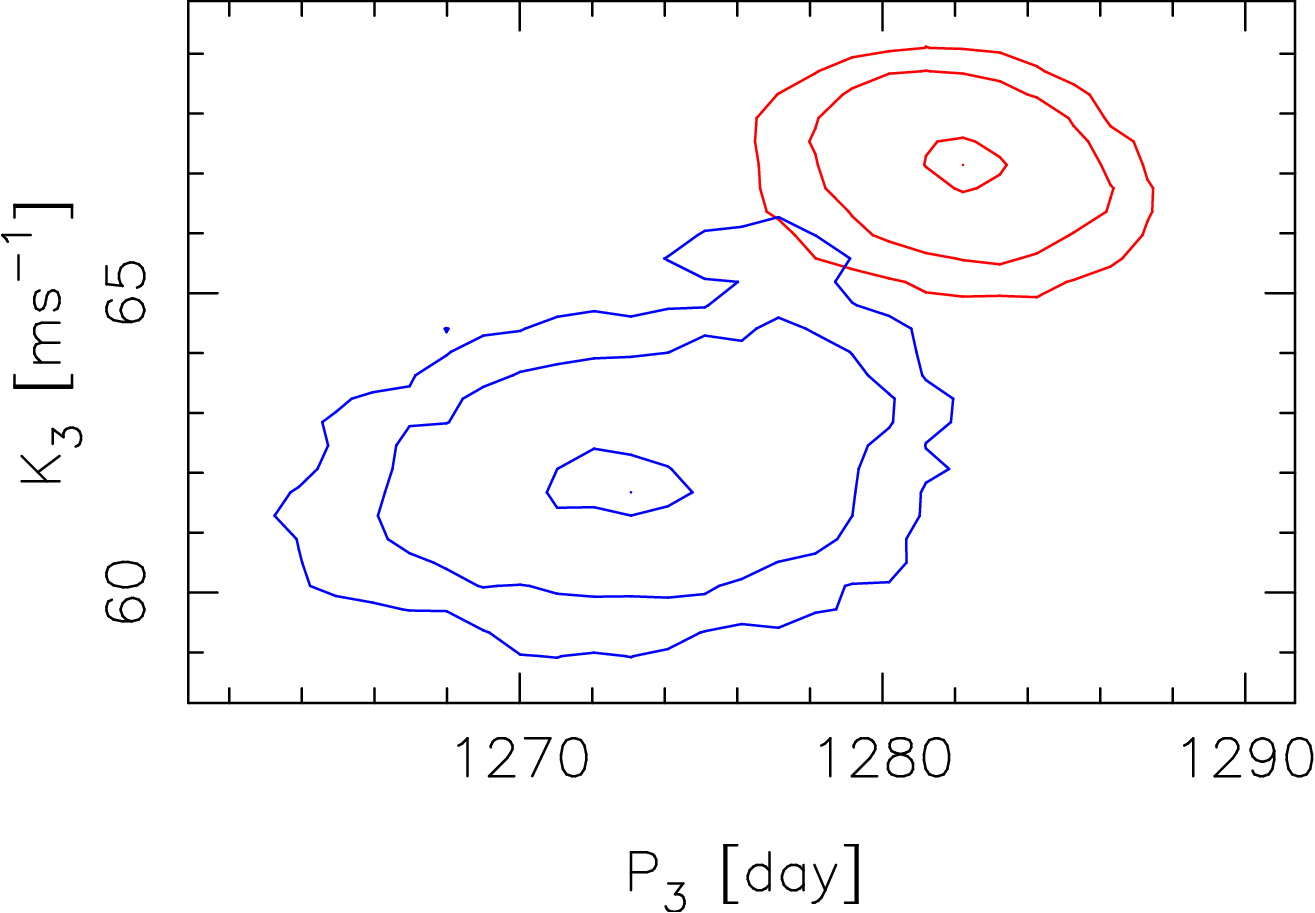}

\includegraphics[angle=270, width=0.45\textwidth]{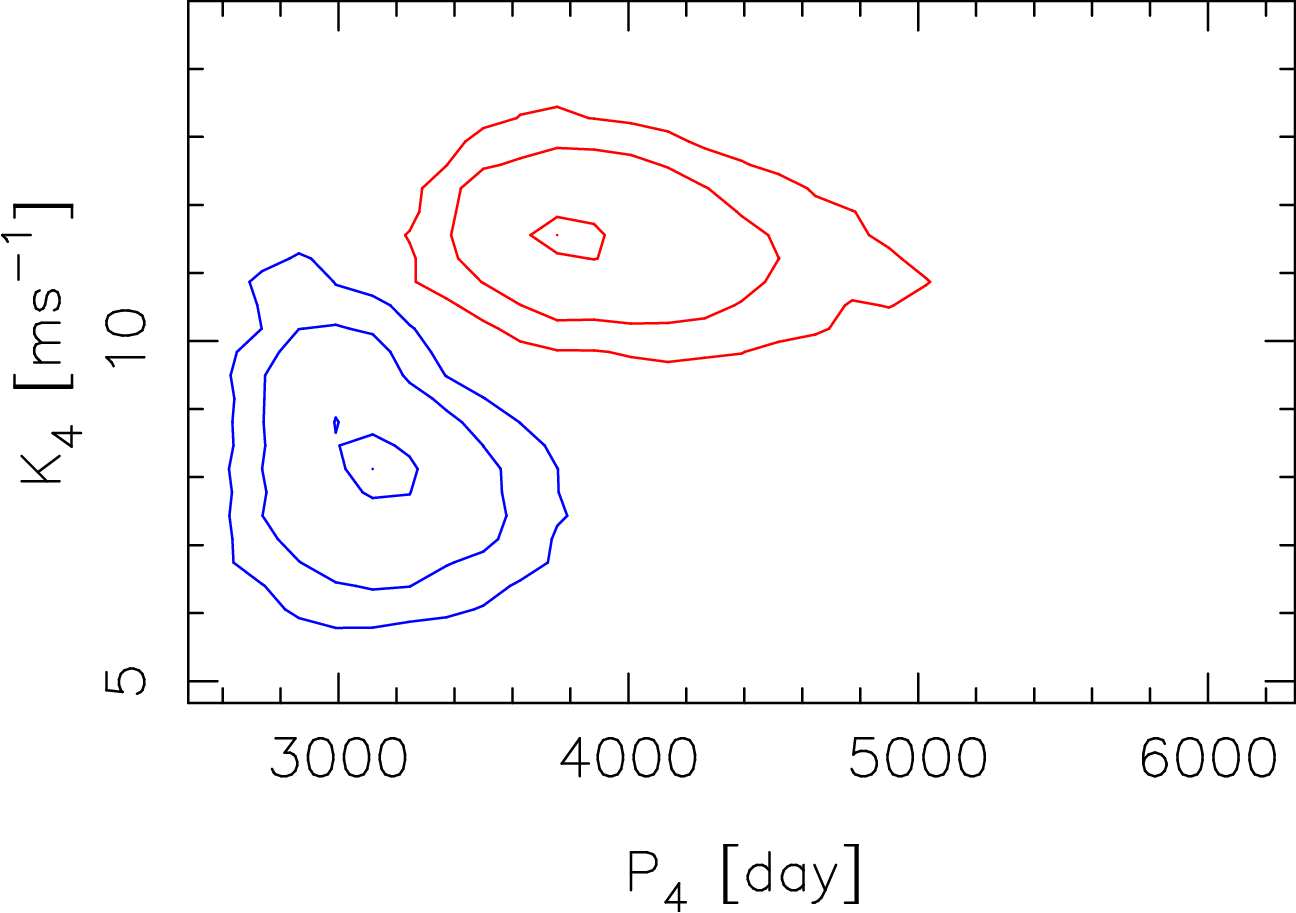}
\caption{The equiprobability contours of the period and amplitude parameters of $\upsilon$ And d (top) and $\upsilon$ And e (bottom) containing 50\%, 95\%, and 99\% of the probability density. The red colour denotes the contours given the Lick1 data of \citet{fischer2003} and blue is used to denote the contours given the Lick2 data of \citet{wright2009}.}\label{upsAnd_contours}
\end{figure}

The Lick1 and Lick2 data sets appear to have one striking difference. While they both imply that there are indeed four Keplerian signals in the $\upsilon$ And RV's, as concluded by \citet{curiel2011}, they do not agree on the orbital period of the proposed fourth signal. The probability of the three companion model is significantly lower than that of the four planet model -- $10^{-4}$ and $10^{-24}$ times lower for Lick1 and Lick2, respectively. This implies that there is either a fourth Keplerian signal in the data or biases that mimic Keplerian periodicity. The MAP estimate and the corresponding $\mathcal{D}_{0.99}$ set of the period of this fourth signal is 3120 [2560, 3940] days for Lick1 data and 3860 [3180, 5160] for Lick2 data. The latter of these estimates appears to be very close to the estimate of \citet{curiel2011} of 3848.86$\pm$0.74 days. However, because of the difference of more than 700 days between the MAP estimates of the periods from Lick1 and Lick2, we cannot conclude, based on the Lick data alone, that there are indeed four Keplerian signals in the data. This inconsistency is seen the most clearly when looking at the equiprobability contours of the parameter posterior densities given each data set. The contours containing 50\%, 95\%, and 99\% of the density are shown in Fig. \ref{upsAnd_contours} for the period and amplitude parameters of $\upsilon$ And d (top) and the proposed $\upsilon$ And e (bottom). The Lick1 contours are shown in red and Lick2 contours in blue. As seen in this Fig., the estimated period and amplitude of the $\upsilon$ And d differ also between the two Lick data sets.

Because of the inconsistency of the Lick data sets published in \citet{fischer2003} and \citet{wright2009}, we use the model inadequacy criterion to find out if either of these two data sets is also inconsistent with the combined ELODIE, AFOE, HET, and HJS data. We denote this combined data as $m$ and use $m_{1}$ and $m_{2}$ to denote the Lick1 and Lick2 data, respectively, and calculate the Bayes factors $B(m,m_{1})$ and $B(m,m_{2})$ for the model $\mathcal{M}_{I,4}$. The logarithms of these factors are 4.01 and -10.20, respectively (Table \ref{dataset_inconsistency}). This implies that the Lick2 data set is inconsistent with the rest of the data and the 4-companion model is an inadequate description with a probability of more than 0.999, whereas the Lick1 data cannot be shown inconsistent with the rest of the data with a probability exceeding 5\%. Therefore, it appears that Lick1 data \citep{fischer2003} is consistent with the other four data sets but the Lick2 data \citep{wright2009} is not. 

We also investigated whether some of the ELODIE, AFOE, HET, HJS, and Lick data sets were inconsistent with the rest of the data by calculating the Bayes factors $B(m_{i}, m)$, where $m_{i}, i=1, ..., 5$, refers to each of these sets, respectively, and $m$ contains all the data except the set $m_{i}$. We performed these calculations using both Lick1 data and Lick2 data. The probabilities of the model $\mathcal{M}_{I,4}$ being inadequate in describing each of these sets with respect the the rest of the data are shown in Table \ref{dataset_inconsistency}.

\begin{table}
\center
\caption{The log-Bayes factors ($\log B$) and probabilities ($P$) of model $\mathcal{M}_{I,4}$ being an inadequate description of each individual set of RV's of $\upsilon$ And and the rest of the data. The Lick1 (L1) and Lick2 (L2) data are analysed separately.\label{dataset_inconsistency}}
\begin{tabular}{lcccc}
\hline \hline
  Set & $\log B$ (L1) & $\log B$ (L2) & $P$ (L1) & $P$ (L2) \\
\hline
  Lick & 4.01 & -10.20 & 0.018 & $>$0.999 \\
  AFOE & -12.91 & -10.81 & $>$0.999 & $>$0.999 \\
  HET & 52.73 & 38.89 & $<10^{-22}$ & $<10^{-16}$ \\
  HJS & 10.55 & 8.70 & $<10^{-4}$ & $<10^{-3}$ \\
  ELODIE & 13.59 & 19.36 & $<10^{-5}$ & $<10^{-8}$ \\
\hline \hline
\end{tabular}
\end{table}

\begin{figure}
\center
\includegraphics[angle=270, width=0.45\textwidth]{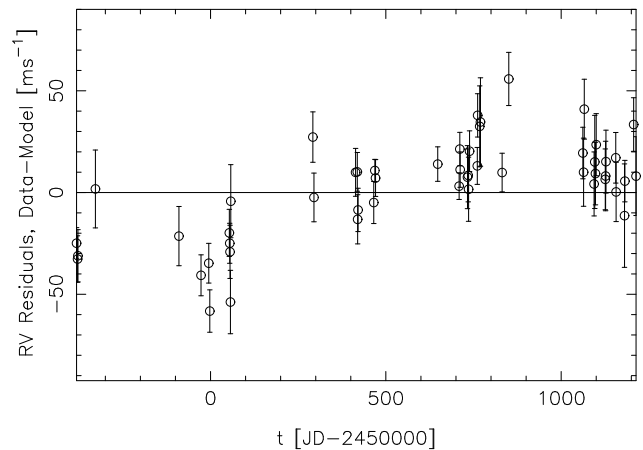}
\caption{The residuals of AFOE RV's of the $\upsilon$ And with the planetary signals subtracted.}\label{AFOE_bias}
\end{figure}

The results in Table \ref{dataset_inconsistency} show that while Lick2 data is inconsistent with the rest of the measurements with respect to the model $\mathcal{M}_{I,4}$, the AFOE data is also inconsistent with the rest of the measurements regardless of using the Lick1 or Lick2 data among the others in the analyses. We also note that the same inconsistency remains for the AFOE data when using the three-companion model $\mathcal{M}_{I,3}$ in the analyses. Therefore, as also noted by \citet{curiel2011}, we conclude that the AFOE data has additional biases and should not be used together with the rest of the data because the results would be prone to biases as well. To further demonstrate the inconsistency of the AFOE data and the other data sets, we show the RV residuals of the AFOE data when the three-companion model has been used to analyse the combined data of AFOE, Lick1, ELODIE, HET, and HJS (Fig. \ref{AFOE_bias}). These residuals appear to show a low-amplitude periodicity that roughly corresponds to the period of companion d, despite the fact that the signal of this companion (and those of b and c) has been subtracted.

We continue the analyses of $\upsilon$ And RV's by neglecting the AFOE data and by using the older Lick1 data set \citep{fischer2003}, because of the inconsistencies of the AFOE and Lick2 data with the rest of the data sets. The combined data set consists of Lick1, HET, ELODIE, and HJS data that contain 248, 79, 71, and 41 measurements, respectively. This combined data set with 439 measurements was analysed using two models, namely, $\mathcal{M}_{I,3}$ and $\mathcal{M}_{I,4}$, because there are clearly three strong Keplerian signals in the data as demonstrated already by \citet{butler1999}, and because the noise levels of the different data sets likely differ from one another based on the previous analyses.

Since we removed the AFOE data from the analyses, we need to assess whether the resulting restricted data set can be shown inadequate or not. For this purpose, we re-calculate the values in Table \ref{dataset_inconsistency} and show them in Table \ref{dataset_inconsistency2}. According to these results, none of the four data sets can be said to conflict with the others. Also, using the Bayesian model inadequacy for multiple data sets by calculating $B(m_{1}, ..., m_{4})$, where $m_{i}, i=1, ..., 4$, correspond to Lick1, ELODIE, HJS, and HET data sets, respectively, we receive a value of 1.4$\times 10^{9}$, which means that these sets are inconsistent with a probability of less than $10^{-9}$ given the four-companion model. Therefore, these four sets can be combined reliably and we calculate our final solution of $\upsilon$ And RV's using these four sets.

\begin{table}
\center
\caption{The log-Bayes factors ($\log B$) and probabilities ($P$) of model $\mathcal{M}_{I,4}$ being an inadequate description of $\upsilon$ And RV's for each individual data set and the rest of the data with the restricted data set of Lick1, HET, HJS, and ELODIE.\label{dataset_inconsistency2}}
\begin{tabular}{lcc}
\hline \hline
  Set & $\log B$ & $P$ \\
\hline
  Lick1 & 23.69 & $<10^{-10}$ \\
  HET & 32.78 & $<10^{-13}$ \\
  HJS & 16.34 & $<10^{-7}$ \\
  ELODIE & 18.06 & $<10^{-7}$ \\
\hline \hline
\end{tabular}
\end{table}

\begin{figure}
\center
\includegraphics[angle=270, width=0.45\textwidth]{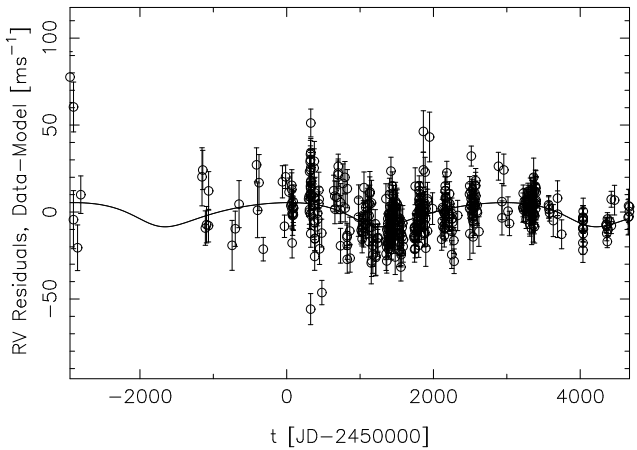}
\caption{The Lick1, ELODIE, HJS, and HET RV's with the signals of the three inner companion removed. The solid curve represents the Keplerian corresponding to planet candidate $\upsilon$ And e.}\label{upsAnd_signal}
\end{figure}

The posterior probability of the model $\mathcal{M}_{I,3}$ is less than $10^{-8}$ of the probability of model $\mathcal{M}_{I,4}$. This implies that there are indeed four periodic signals in the combined data set. The revised orbital parameters with respect to the four-companions model are shown in Table \ref{upsAnd_parameters}. The RV variations corresponding to the longest periodicity in the data are shown in Fig. \ref{upsAnd_signal} together with the fitted Keplerian signal. The signals of the three inner companions have been subtracted from the residuals in Fig. \ref{upsAnd_signal}.

\begin{table*}
\center
\caption{The four-planet solution of $\upsilon$ Andromedae RV's from Lick1, HET, ELODIE, and HJS. MAP estimates of the parameters and the limits of their $\mathcal{D}_{0.99}$ sets.\label{upsAnd_parameters}}
\begin{tabular}{lcccc}
\hline \hline
Parameter & Planet b & Planet c & Planet d & Planet e \\
\hline
$P$ [days] & 4.617098 [4.617047, 4.617174] & 241.50 [241.31, 241.70] & 1278.4 [1271.2, 1285.6] & 2860 [2600, 3220] \\
$e$ & 0.022 [0, 0.047] & 0.278 [0.250, 0.311] & 0.307 [0.272, 0.339] & 0.13 [0, 0.28] \\
$K$ [ms$^{-1}$] & 71.0 [69.0, 72.7] & 52.8 [51.0, 55.2] & 61.6 [59.1, 64.3] & 7.1 [4.9, 9.4] \\
$\omega$ [rad] & 1.4 [0.0, 3.0] & 4.15 [3.99, 4.30] & 4.46 [4.32, 4.62] & 2.6 [0.2, 5.1] \\
$M_{0}$ [rad]& 2.8 [1.4, 4.4] & 3.97 [3.82, 4.11] & 0.29 [0.17, 0.41] & 2.4 [0.0, 5.1] \\
$m_{p} \sin i$ [M$_{J}$] & 0.683 [0.617, 0.748] & 1.91 [1.70, 2.09] & 3.85 [3.47, 4.28] & 0.58 [0.40, 0.78] \\
$a$ [AU] & 0.0589 [0.0560, 0.0615] & 0.823 [0.783, 0.860] & 2.50 [2.38, 2.62] & 4.27 [3.95, 4.66] \\
\hline
$\gamma_{1}$ [ms$^{-1}$] (Lick) & 3.7 [1.3, 6.0] \\
$\gamma_{2}$ [ms$^{-1}$] (ELODIE) & -12.7 [-15.9, -7.9] \\
$\gamma_{3}$ [ms$^{-1}$] (HJS) & -15.4 [-19.5, -10.8] \\
$\gamma_{4}$ [ms$^{-1}$] (HET) & -19.4 [-22.2, -16.8] \\
$\sigma_{I,1}$ [ms$^{-1}$] (Lick) & 7.68 [5.89, 9.47] \\
$\sigma_{I,2}$ [ms$^{-1}$] (ELODIE) & 16.3 [12.2, 20.4] \\
$\sigma_{I,3}$ [ms$^{-1}$] (HJS) & 8.6 [1.5, 18.8] \\
$\sigma_{I,4}$ [ms$^{-1}$] (HET) & 1.95 [0, 4.58] \\
\hline \hline
\end{tabular}
\end{table*}

When comparing the orbital parameters of our solution in Table \ref{upsAnd_parameters} with the solution of \citet{curiel2011}, it can be seen that the period of the $\upsilon$ And e is significantly lower in our solution. We received a MAP estimate for the orbital period of 2860 days ($\mathcal{D}_{0.99} = [2600, 3220]$), whereas \citet{curiel2011} reported a period of 3848.86$\pm$0.74 days. This difference can arise from the fact that they used the more recent Lick2 data set of \citet{wright2009}, which is not consistent with the other RV's according to our analyses. We also found another solution for the period of $\upsilon$ And e. This period is 5750 days ($\mathcal{D}_{0.99} = [5220, 6610]$), roughly twice the MAP periodicity, but its posterior probability is more than a thousand times lower than that of the solution in Table \ref{upsAnd_parameters}.

We note that while \citet{curiel2011} adopted a jitter of 10 ms$^{-1}$ when analysing the RV's of $\upsilon$ And, the estimate of \citet{butler2006} is only 4.2 ms$^{-1}$. Our results are consistent with the latter estimate because the upper limit of excess noise, including the stellar jitter, is 4.58 ms$^{-1}$ based on the lowest noise level in the data sets of the HET data (Table \ref{upsAnd_parameters}). According to our results, the jitter has likely an even lower value of roughly 2.0 ms$^{-1}$. This also implies that the Lick1, ELODIE, and HJS data contain an additional source of RV variations -- likely the telescope-instrument combination used to measure these data.

\section{Discussion}

We have proposed a simple method for assessing whether a statistical model is an inadequate description of multiple independent data sets. This method is simply an application of the well known Bayesian model selection theory and the law of conditional probability but it also differs from the common model selection approach because it provides the means of determining whether a single model, i.e. the best model in the \emph{a priori} selected model set, is not an adequate description of the data sets and needs to be improved.

Using this Bayesian model inadequacy criterion and common model comparisons, we re-analysed three combined RV data sets made using at least two telescope-instrument combinations. According to our results, the Gliese 581 RV's observed using the HIRES and HARPS spectrographs can be described reliably using the model $\mathcal{M}_{I,4}$, where their uncertainties caused by stellar jitter and additional instrument uncertainty have been modelled to have different magnitudes -- at least, the four-companion model cannot be shown to be an inadequate description of these two data sets. This suggests that the results in \citet{tuomi2011} are indeed reliable in this respect.

The RV's of HD 207107 showed that there can be significant telescope-instrument -induced uncertainties in the data. Therefore, we were forced to describe these uncertainties with different parameters for each telescope-instrument -combination. According to our results, the telescope-instrument uncertainties can differ considerably between different data sets, which makes it more difficult to put reliable constraints to the stellar jitter. While the jitter of HD 217107 is not likely to exceed 6.0 ms$^{-1}$ based on the noise in the Euler data, the Lick1 data turned out to have excess noise of 5-10 ms$^{-1}$ with respect to this jitter estimate (Table \ref{HD217107_parameters}). Therefore, we conclude that the instrument uncertainties cannot be assumed as known, and additional noise should always be assumed to exist in the data. When neglecting this additional uncertainty, the estimates of orbital parameters can be biased and their uncertainty estimates will certainly be unrealistically low with respect to the information in the measurements.

The RV's of $\upsilon$ Andromedae proved a challenging analysis problem on their own. These data consisted of five independent RV data sets. According to our results, the Lick2 data of \citet{wright2009} was not consistent with the other data sets with respect to our model inadequacy criterion but the earlier Lick1 data set of \citet{fischer2003} should be used instead. Unfortunately, it is not possible to tell where this inadequacy arises from. Also, the AFOE data \citep{butler1999} turned out to contradict with the rest of the data, likely because of biases in the process of making the measurements, as also noted by \citet{curiel2011}. This leaved only four consistent data sets, Lick1 \citep{fischer2003}, ELODIE \citep{naef2004}, HET \citep{mcarthur2010}, and HJS \citep{wittenmyer2007}, to be used in the analyses. With differing noise levels for each of these sets, we calculated the revised orbital parameters for the $\upsilon$ And planetary system with four planetary companions (Table \ref{upsAnd_parameters}).

Because our four-planet solution of the RV's of $\upsilon$ And differs significantly from the proposed solution of \citet{curiel2011} with respect to the orbital period of the outer planet, numerical integrations of the orbits are needed to assess the stability of our solution. The lower estimate for the orbital period of $\upsilon$ And e does not support the conclusion that the d and e planets could be in a 3:1 mean motion resonance (MMR). However, our solution coincides roughly with a 2:1 MMR, which could enable the stability of the system over long time-scales. Investigating the stability of our solution is necessary to be able to determine whether it corresponds to a physically viable system and is not simply an artefact caused by noise, data sampling, and possible biases in the measurements.

For successful detections, it is crucial that the noise -- i.e. all the other variations except the Keplerian signals -- in the valuable measurements is modelled as realistically as possible and not simply minimised as is commonly the case when using simple $\chi^{2}$ minimisations and related methods. Our method can be used readily to detect whether the statistical model indeed describes the data adequately with respect to the selected noise model as well.

The application of our criterion to measurements of any complex systems is obvious. Systems whose behaviour, time-evolution, and dependence on different physical and other factors cannot be derived from fundamental physical principles, are difficult to model because the models are necessarily empirical descriptions, whose validity can only be assessed using measurements. In such systems, there can be numerous small-scale effects and/or biases, whose existence is not known and whose magnitude cannot be measured. These effects cannot therefore be taken into account in the model constructed to describe some desired features of the system. As briefly noted in \citet{kaasalainen2011}, the ability to show that a model is an insufficient description of the measurements is therefore needed to be able to determine whether the model needs to be improved further to extract all the valuable information from the noisy data. According to the demonstrations in this article, our method can be said to satisfy these needs to significant extent. Also, as we did not make any assumptions regarding the exact nature of the model, the criterion can be applied to any problem for which it is possible to calculate the likelihoods of the measurements using the model.

Finally, we note that if the model has been constructed prior to the measurements, the model inadequacy means that the earlier data sets used to construct the model, i.e. to select the model formulae and calculate the posterior densities of the model parameters, conflict with the new ones with respect to the model. It could also be that the model is being developed using a single data set in hand. Then, despite being the best model in the sense of having the greatest posterior probability, the model could still be inadequate in describing some part of the data set with respect to another part \citep{kaasalainen2011}. Either way, the measurements cannot be described adequately using the selected model and we say that the model is inadequate. Our criterion can be used in these cases as well.

\begin{acknowledgements}
M. Tuomi is supported by RoPACS (Rocky Planets Around Cools Stars), a Marie Curie Initial Training Network funded by the European Commission's Seventh Framework Programme.
\end{acknowledgements}

\begin{appendix}

\section{Model inadequacy criterion}

\subsection{Two data sets}

We start by defining what we mean by model inadequacy in describing two sets of data and derive its equations from the common Bayesian model comparison theory.

We assume that there are independent measurements, or series of measurements, $m_{i} : i=1, ..., N, N \geq 2$, that have been made to study the same system of interest. Because these measurements describe the same system, or at least contain information on the same aspects of the system of interest, they can be modelled with statistical models that have at least one parameter in common, namely, $\theta \in \Theta$. Throughout this article the parameter space $\Theta$ is a bounded subset of $\mathbb{R}^{k}$. The parameter $\theta$ is used to quantify some features in the measurements $m_{i}, \forall i$. In addition, there are other parameters, namely $\omega_{i} \in \Theta : i = 1, ..., N$, that each quantify some additional features in the $i$th measurement.

The measurements can now be used to compare different statistical models using the Bayesian model selction theory. Let $P(\mathcal{A} | m_{i}, m_{j})$ be the posterior probability of model $\mathcal{A}$ given the measurements $m_{i}$ and $m_{j}$. The model $\mathcal{A}$ can be any model for which a likelihood function exists. With this model, both measurements are modelled using the same parameter $\theta$ and different parameters $\omega_{i}$ and $\omega_{j}$, respectively. Probability $P(\mathcal{B} | m_{i}, m_{j})$ is the corresponding probability when measurements $m_{i}$ and $m_{j}$ are modelled using the same model structure as model $\mathcal{A}$ has, but this time with parameters $\phi_{i}$ and $\phi_{j}$, where $\phi_{k} = (\theta_{k}, \omega_{k}), k = i,j$, respectively. 

Therefore, because of the independence of the measurements and the independence of $\phi_{i}$ and $\phi_{j}$, the marginal integral in Eq. (\ref{marginal_likelihood}) of the measurements with respect to the model $\mathcal{B}$ can be written as\footnote{The reader should refer to any basic text on conditional probabilities and independence.}
\begin{eqnarray}\label{model_likelihood}
  && P(m_{i}, m_{j} | \mathcal{B}) \nonumber\\
  && = \int_{\phi_{i}, \phi_{j} \in \Theta} l(m_{i}, m_{j} | \phi_{i}, \phi_{j}, \mathcal{B}) \pi(\phi_{i}, \phi_{j} | \mathcal{B}) d (\phi_{i}, \phi_{j}) \nonumber\\
  && = \prod_{k=i,j} \int_{\phi_{k} \in \Theta} l(m_{k} | \phi_{k}, \mathcal{B}) \pi(\phi_{k} | \mathcal{B}) d \phi_{k} = P(m_{i} | \mathcal{A}) P(m_{j} | \mathcal{A}) ,
\end{eqnarray}
where the model has been changed to $\mathcal{A}$, because given only one measurement, $\mathcal{A}$ and $\mathcal{B}$ are in fact the same model. In the above, we have used $l$ and $\pi$ to denote the likelihood function and the prior density, respectively.

Now, let $s \in [0,1]$ be a small threshold probability. We compare the probabilities of the models $\mathcal{A}$ and $\mathcal{B}$ given the measurements $m_{i}$ and $m_{j}$. If $P(\mathcal{A} | m_{i}, m_{j}) < s$, we say, that the model is an inadequate description of the data with a probability of $1-s$ and that the model $\mathcal{A}$ cannot be used to model them both. In other words, the probability of model $\mathcal{A}$ is so small, that the measurements should instead be modelled using different parameters $\theta_{i}$ and $\theta_{j}$, i.e. using model $\mathcal{B}$. This condition is simply the common Bayesian model selection criterion \citep[e.g.][]{jeffreys1961}. From this condition and Eq. (\ref{model_likelihood}), and when selecting the prior probabilities of the two models equal, the comparison of models $\mathcal{A}$ and $\mathcal{B}$ according to Eq. (\ref{model_probability}) leads to
\begin{equation}\label{model_probabilities}
  P(m_{i}, m_{j} | \mathcal{A}) < \frac{s}{1-s} P(m_{i} | \mathcal{A}) P(m_{j} | \mathcal{A}) .
\end{equation}
We denote $r = s(1-s)^{-1}$ and leave the model out of the notation by denoting $P(m) = P(m | \mathcal{A})$ when it is clear which model has been used. Now, we define the model inadequacy as follows.

The model used to describe the measurements $m_{i}$ and $m_{j}$ is not adequate with level $r$ if
\begin{equation}\label{inadequacy_defn}
  B(m_{i}, m_{j}) := \frac{P(m_{i}, m_{j})}{P(m_{i}) P(m_{j})} < r ,
\end{equation}
where the factor $B$ is actually the Bayes factor in favour of model $\mathcal{A}$ and against model $\mathcal{B}$ and $r$ is some (small) positive number corresponding to the selected threshold probability $s$.

Because we have made no assumption on the exact nature of the measurements, the model, or the modelled system, the above condition applies to anything that can be measured and described with a statistical model. In fact, to be able to use the Eq. (\ref{inadequacy_defn}), a sufficient condition is that the measurements $m_{i}$ and $m_{j}$ are modelled using statistical models that have at least one parameter, namely $\theta$, in common. The model of the $i$th data set may have other parameters $\omega_{i}$ and these have to be treated as free parameters as well, but they have no role in the Eq. (\ref{inadequacy_defn}) because they are independent of the other data set.

The Eq. (\ref{inadequacy_defn}) in fact states that the measurements are not distributed according to the model used. However, the converse is not true. If the condition in Eq. (\ref{inadequacy_defn}) does not hold for some measurements $m_{i}$ and $m_{j}$, it cannot be said that they are drawn from the same modelled density, even though it might be a reasonable assumption in practice.


The Bayes factor in Eq. (\ref{inadequacy_defn}) has an interesting property when interpreted in terms of the information gain defined using the Kullback-Leibler (K-L) divergence \citep{kullback1951} between prior and the posterior. The K-L divergence is defined for two continuous random variables with probability densities $u(x)$ and $v(x)$ as
\begin{equation}\label{KLD}
  D_{KL}\big\{u(x) || v(x)\big\} = \int u(x) \log \frac{u(x)}{v(x)} dx .
\end{equation}
With this notation, we can write the K-L divergence of moving from the prior to the posterior (given both data sets). Hence, it follows that
\begin{eqnarray}\label{gain_inadequacy}
  && D_{KL}\big\{\pi(\theta | m_{i}, m_{j}) || \pi(\theta)\big\} = \int \pi(\theta | m_{i}, m_{j}) \log \frac{\pi(\theta | m_{i}, m_{j})}{\pi(\theta)} d \theta \nonumber\\
  && = - \log P(m_{i}, m_{j}) + \int \pi(\theta | m_{i}, m_{j}) \log l(m_{i}, m_{j} | \theta) d \theta \nonumber\\
  && = -\log P(m_{i}, m_{j})  + \sum_{k=i,j} \int \pi(\theta | m_{i}, m_{j}) \log \frac{l(m_{i}, m_{j} | \theta)}{l(m_{k} | \theta)} d \theta \nonumber\\
  && \Leftrightarrow \log B(m_{i}, m_{j}) = D_{KL} \big\{ \pi(\theta | m_{i}, m_{j}) || \pi(\theta) \big\} \nonumber\\
  && - D_{KL} \big\{ \pi(\theta | m_{i}, m_{j}) || \pi(\theta | m_{i}) \big\} - D_{KL} \big\{ \pi(\theta | m_{i}, m_{j}) || \pi(\theta | m_{j}) \big\} ,
\end{eqnarray}
where we have used the Bayes rule and the facts that integral over a probability density equals unity and $m_{i}$ and $m_{j}$ are independent.

This means that the logarithm of the Bayes factor used to determine the model inadequacy in describing measurements $m_{i}$ and $m_{j}$ can in fact be interpreted as the total information gain of the two measurements minus the information gains of moving from the posterior with respect to each measurement alone to the full posterior.

Alternatively, the Bayes factor can be written using the information losses, or K-L divergences, of moving from the posteriors back to the prior (as opposed to the information gain of moving from prior to the posterior). With this terminology, and using a similar derivation as for the information gain in Eq. (\ref{gain_inadequacy}), the expression in Eq. (\ref{gain_inadequacy}) can be replaced by
\begin{eqnarray}\label{loss_inadequacy}
  && \log B(m_{i}, m_{j}) = D_{KL} \big\{\pi(\theta) || \pi(\theta | m_{i}, m_{j}) \big\} \\
  && - D_{KL} \big\{\pi(\theta) || \pi(\theta | m_{i}) \big\} - D_{KL} \big\{ \pi(\theta) || \pi(\theta | m_{j}) \big\} , \nonumber
\end{eqnarray}
which means that the logarithm of the Bayes factor can be interpreted as the total information loss of the two measurements minus the information losses of the two measurements separately.

\subsection{Multiple data sets}

When there are more than two data sets available, the model inadequacy criterion can be derived easily following the considerations in the previous subsection. For measurements $m_{i}, i = 1, ..., N$, it can be seen that
\begin{equation}\label{multiple_likelihood}
 P(m_{1}, ..., m_{N} | \mathcal{B}) = \prod_{i=1}^{N} P(m_{i} | \mathcal{A}) .
\end{equation}
It then follows that the model inadequacy criterion corresponding to that in Eq. (\ref{model_probabilities}) can be written as
\begin{equation}\label{multiple_probabilities}
 P(m_{1}, ..., m_{N} | \mathcal{A}) < \frac{s}{1-s} \prod_{i=1}^{N} P(m_{i} | \mathcal{A}) .
\end{equation}
We again use $B$ to denote the Bayes factor and write this criterion in the following way.

The model used to describe measurements $m_{i}, ..., m_{N}$ does not describe the measurements adequately accurately with level $r$ if
\begin{equation}\label{multiple_defn}
  B(m_{1}, ..., m_{N}) : = \frac{P(m_{1}, ..., m_{N})}{\prod_{i} P(m_{i})} < r .
\end{equation} 

From the Eq. (\ref{multiple_defn}), it can be seen that for $N$ data sets, the marginal integral needs to be determined $N+1$ times to receive the Bayes factor that is used to assess the model inadequacy. This requirement cannot be considered very limiting, because in practice, the data sets are commonly analysed separately anyway.

In terms of K-L information loss of moving from the posterior to the prior, the Bayes factor $B$ can again be interpreted in a simple manner using similar derivation as in Eq. (\ref{gain_inadequacy}). As a consequence, it follows that
\begin{eqnarray}\label{multiple_loss}
  && \log B(m_{1}, ..., m_{N}) = D_{KL} \big\{ \pi(\theta) || \pi(\theta | m_{1}, ..., m_{N}) \big\} \nonumber\\
  && - \sum_{i=1}^{N} D_{KL} \big\{ \pi(\theta) || \pi(\theta | m_{i}) \big\} .
\end{eqnarray}
However, the information gains cannot be used in a similar manner as in Eq. (\ref{gain_inadequacy}). Instead, using the information gain of the measurements the generalisation of Eq. (\ref{gain_inadequacy}) to several measurements is
\begin{eqnarray}\label{multiple_gain}
  && \log \frac{\prod_{i} B(m_{i}, (m_{1}, ..., m_{k}, ..., m_{N}) |_{k \neq i})}{B(m_{1}, ..., m_{N})} \nonumber\\
  && = D_{KL} \big\{ \pi(\theta | m_{1}, ..., m_{N}) || \pi(\theta) \big\} \nonumber\\
  && - \sum_{i=1}^{N} D_{KL} \big\{ \pi(\theta | m_{1}, ..., m_{N}) || \pi(\theta | m_{1}, ..., m_{k}, ..., m_{N}) |_{k \neq i} \big\} ,
\end{eqnarray}
where $B(m_{i}, (m_{1}, ..., m_{k}, ..., m_{N})|_{k \neq i})$ is the Bayes factor describing the model inadequacy with respect to two data sets, namely, $m_{i}$ and the combined data set $(m_{1}, ..., m_{k}, ..., m_{N})|_{k \neq i}$, which denotes all the data except the measurement $m_{i}$.

Therefore, the Bayes factor determining the model inadequacy in Eq. (\ref{multiple_defn}) can be interpreted as a measure of information loss that results from disregarding the measurements to gain information on the posterior minus the corresponding information losses of disregarding each measurement one at the time. Naturally, the gain and loss Eqs. (\ref{multiple_gain}) and (\ref{multiple_loss}) are equivalent if $N = 2$, as was seen in the previous subsection.

Assuming that $B(m_{1}, ..., m_{N}) \geq 1$, which means that model $\mathcal{A}$ has a greater probability than $\mathcal{B}$, has an interesting consequence. From this assumption, it follows that
\begin{eqnarray}\label{information_gain}
  && D_{KL} \big\{ \pi(\theta) || \pi(\theta | m_{1}, ..., m_{N}) \big\} \geq \sum_{i=1}^{N} D_{KL} \big\{ \pi(\theta) || \pi(\theta | m_{i}) \big\} .
\end{eqnarray}
When again interpreted in terms of information loss, this means that given a model that cannot be shown inadequate with $r = 1$, the amount of information in the combined data set is greater than the information in the individual data sets.

\end{appendix}

\end{document}